\documentclass{article}

\usepackage{arxiv} 

\usepackage[utf8]{inputenc} 
\usepackage[T1]{fontenc}    
\usepackage[hidelinks]{hyperref}       
\usepackage{url}            
\usepackage{booktabs}       
\usepackage{amsfonts}       
\usepackage{nicefrac}       
\usepackage{microtype}      
\usepackage{lipsum}
\usepackage{amsmath}

\usepackage{xcolor}
\usepackage{soul}

\usepackage{graphicx}
\graphicspath{{../../lpde/fig/paper/}}
\graphicspath{{fig/}}

\title{Learning emergent PDEs  \\
in a learned emergent space} 

\author{%
  Felix P.~Kemeth\\
  Department of Chemical and Biomolecular Engineering\\
  Whiting School of Engineering, Johns Hopkins University\\
  3400 North Charles Street, Baltimore, MD 21218, USA\\
  \And
  Tom Bertalan \\
  Department of Chemical and Biomolecular Engineering\\
  Whiting School of Engineering, Johns Hopkins University\\
  3400 North Charles Street, Baltimore, MD 21218, USA\\
  \And
  Thomas Thiem\\
  The Department of Chemical and Biological Engineering \\
  Princeton University\\
  Princeton, NJ 08544, USA\\
  \And
  Felix Dietrich\\
  Institut f\"{u}r Informatik, TU M\"{u}nchen\\
  Boltzmannstr. 3\\
  85748 Garching b. M\"{u}nchen, Germany\\
  \And
  Sung Joon Moon\\
  The Department of Chemical and Biological Engineering \\
  Princeton University\\
  Princeton, NJ 08544, USA\\
  \And
  Carlo R. Laing \\
  School of Natural and Computational Sciences\\
  Massey University (Albany)\\
  Private Bag 102-904, Auckland New Zealand \\
  \And
  Ioannis G. Kevrekidis\\
  Department of Chemical and Biomolecular Engineering\\
  Whiting School of Engineering, Johns Hopkins University\\
  3400 North Charles Street, Baltimore, MD 21218, USA\\
  \texttt{yannisk@jhu.edu}
}

\date{\today}

\begin{document}

\maketitle

\begin{abstract}
  We extract data-driven, intrinsic spatial coordinates from observations of the dynamics of large systems of coupled heterogeneous agents.
  These coordinates then serve as an {\em emergent space} in which to learn predictive models in the form of partial differential equations (PDEs) for the collective description of the coupled-agent system.
  They play the role of the independent spatial variables in this PDE (as opposed to the dependent, possibly also data-driven, state variables).
  This leads to an alternative description of the dynamics, local in these emergent coordinates, thus facilitating an alternative modeling path for complex coupled-agent systems.
  %
  We illustrate this approach on a system where each agent is a limit cycle oscillator (a so-called Stuart-Landau oscillator); the agents are heterogeneous (they each have a different intrinsic frequency $\omega$) and are 
  coupled through the ensemble average of their respective variables.
  After fast initial transients, we show that the collective dynamics on a slow manifold can be approximated through a learned model based on local ``spatial'' partial derivatives in the emergent coordinates.
  The model is then used for prediction in time, as well as to capture collective bifurcations when system parameters vary.
  The proposed approach thus integrates the automatic, data-driven extraction of
  \textit{emergent space} coordinates parametrizing the agent dynamics, with machine-learning assisted identification of an ``emergent PDE'' description of the dynamics in this parametrization.
  %
  %
  %
  %
  %
\end{abstract}


\section{Introduction}
\label{sec:int}

Modeling the dynamic behavior of large systems of interacting agents remains a challenging problem in complex systems analysis.
Due to the large state space dimension of such systems, it has historically been an ongoing research goal to construct useful reduced-order models with which to collectively
describe the coarse-grained dynamics of agent ensembles.
Such coarse-grained, collective descriptions arise in many contexts, e.g. in  thermodynamics, where interacting particles may effectively be described at the macroscopic level by temperature, pressure and density; or in kinetic theory, where collisions in the Boltzmann equation can lead to 
continuum descriptions, such as the Navier-Stokes equations - but also in
contexts such as chemotaxis or granular flows.
One important issue in this coarse-graining is to find coarse-grained observables (density fields, momentum fields, concentration fields, void fraction fields) that describe the collective behavior in physical space.
Macroscopic, effective models are then often approximated as partial differential equations (PDEs) for these fields: their time derivatives are expressed locally in terms of the local spatial derivatives of the field(s) at each point.
The closures required to derive predictive models can be obtained either mathematically (with appropriate assumptions) and/or semi-empirically through experimental or computational observations. 

When the interacting agents are coupled oscillator systems, their observed low-dimensional dynamics can sometimes be described as a ``lumped'' system of a few ordinary differential equations (ODEs) in terms of 
so-called {\em order parameters}~\cite{kuramoto1984_chemical_osci, STROGATZ20001, Ott-Antonsen}.
For large heterogeneous systems of interacting oscillators we observe, at any given moment, a distribution of oscillator states; being able to usefully describe this evolution by a few ODEs for appropriate order parameters corresponds, conceptually, to describing the distribution evolution through a finite, closed set of a few moment equations for the distribution. The ``few good'' order parameters are here provided by the few leading moments in terms of which a closed set of model ODEs (or even stochastic differential equations) can be written.
And while in some cases such a reduced description can be quite successful, there are other cases where a few ODEs will not suffice, and where one needs to write evolution equations (e.g. PDEs) for evolving {\em field(s)} of instantaneous oscillator behavior(s).

The question then naturally arises: What is a good way of parametrizing the spatial support of this evolving distribution of behaviors? Which (and how many) are the few {\em independent}, ``spatial'' variables, in the space of which we will attempt to derive evolutionary PDE models for the collective behavior evolution? 
In other words, when the problem does not evolve in physical space (e.g. when the oscillators are nodes in an interacting network) {\em does there exist} a useful continuum embedding space in which we can observe the behavior evolving as a spatiotemporal field? And if so, how can we detect this {\em emergent space} and its parametrizing independent coordinates in a data-driven way, based on observations of the collection of individual coupled agent dynamics? 
%
%
Our task thus has two components, both accomplished here in a data-driven fashion:
(a) find emergent ``spatial'' coordinates in which the oscillator behavior can be (embedded and) observed as smooth
spatiotemporal field evolution; and
(b) once these emergent coordinates have been obtained, learn a model of the evolving dynamics,
if possible in the form of a partial differential equation governing this field; that is, approximate
the (pointwise) time derivative(s) of the field(s) in terms of a few local spatial derivatives of the 
field {\em in the emergent independent variables}. 

The data-driven approximation of such evolution operators for spatiotemporal dynamics using machine learning tools (neural networks, Gaussian processes, manifold learning....) is a long-standing research endeavor - we, among others, have worked on neural network-based identification of nonlinear distributed  systems~\cite{krischer93_model_ident_spatiot_varyin_catal_react,rico-martinez92_discr_vs, gonzalez-garcia98_ident_distr_param_system}; the subject is currently exploding in the machine learning literature, e.g. ~\cite{brunton20_machin_learn_fluid_mechan, lu2020deeponet}.
The ``twist'' in our work here is that the space in which the evolution operator
(that is, the PDE)  will be learned (the independent variables in which the ``spatial derivatives'' will be estimated) is not known {\em a priori} but will be rather identified, in a first step, through data mining/manifold learning~\cite{kemeth18_emerg_space_distr_data_with, arbabi2020coarsegrained}. 
If/when such an approach is successful, it can lead to a dramatic reduction of the computational cost
of simulation/prediction of the collective, coarse-grained dynamics (compared to the individual evolution of every oscillator/agent in the ensemble). 
This reduced description also enables tasks (effective stability and bifurcation analysis, even control and optimization) that would be difficult or impossible to perform with the fine-scale model.
More importantly, if successful and generalizable enough, this alternative description in terms of field PDEs in emergent variables, (assisted by computationally mapping back-and-forth between fine and coarse descriptions) may guide a new, coarse-grained interpretation and even understanding of the system dynamics. 

There may appear to be a contradiction between having fine-scale dynamics we know to involve long-range interactions (here, all-to-all coupling), and then learning a model based on local interactions (here, coupling with oscillators that have nearby behavior, through local ``behavior derivatives'' in our emergent space). We will return to this issue repeatedly in the discussion below, but we mention that the learned operators are not themselves ``the true physics''; they are but a particular, parsimonious parametrization of the long-term dynamics (after initial transients) on a much lower-dimensional slow manifold on which the collective behavior evolves. It is the low dimensionality of this manifold, and the power of embedding theorems like those of Whitney~\cite{Whitney} and Takens~\cite{Takens1981} that enable data-driven {\em parameterizations} (as opposed to physically meaningful mechanistic interpretations) of the long-term dynamics. The many coupled local grid points underpinning a finite-difference discretization of a PDE will here play the role of the many ``generic observers'' parametrizing the relatively low-dimensional manifold on which the coarse-grained long-term dynamics and the attractors of the system are expected to live. 

This approach is fundamentally different from recent approaches where the dynamics are learned in a latent space of {\em dependent variables}, typically as systems of ODEs (but also PDEs with {\em known} independent variables).
Examples of these ``dependent variable latent spaces'' include learning the dynamics of spatial principal component coefficients on an inertial
manifold~\cite{linot20_deep_learn_to_discov_predic}
or learning an ODE in a latent space of an autoencoder using dictionaries
and sparsity promoting regularization~\cite{champion19_data_driven_discov_coord_gover_equat}.
Since early works (e.g. see~\cite{lapedes} on the Mackey-Glass equation, also Refs.~\cite{hudson,rico-martinez92_discr_vs,gonzalez-garcia98_ident_distr_param_system}),
%
learning dynamical systems from data has regained increased attention in recent years.
Popular examples include (in a vast literature) sparse identification of nonlinear dynamical systems using dictionaries~\cite{brunton16_discov_gover_equat_from_data}, DeepXDE\cite{lu19_deepx}, neural ODEs~\cite{duvaneau},
LSTM neural networks~\cite{vlachas18_data_driven_forec_high_dimen} and PDE-net~\cite{long17_pde_net}.
As in the latter, the emergent PDE will be learned here from discrete time data using an explicit forward Euler time integration step (in effect, training a ResNet); many other approaches are also possible (for a ResNet-like Runge-Kutta recurrent network, see Ref.~\cite{gonzalez-garcia98_ident_distr_param_system}).
%

To find coordinates in which to learn the PDE description, we follow the recent work~\cite{kemeth18_emerg_space_distr_data_with,thiem20_emergent_spaces_for_coupled_oscillators} and use diffusion maps~\cite{Nadler2006,Coifman2006},
a nonlinear manifold learning technique.
As our agent-based example, we use coupled Stuart-Landau oscillators,
\begin{equation}
  \frac{\mathrm{d}}{\mathrm{d}t} W_k = \left(1+i \omega_k\right) W_k - \left|W_k\right|^2W_k  + \frac{K}{N}\sum_{j=1}^N\left( W_j - W_k\right)\label{eq:sle};
\end{equation}
each oscillator $k=1,\dots, N$ is represented by a complex variable
$W_k$ and coupled to all other oscillators through the ensemble average.
The long-range interaction is in fact global, since the coupling is all-to-all.
Each agent, when uncoupled, undergoes periodic motion with its own intrinsic frequency $\omega_k$, different across agents, making the ensemble heterogeneous.

\begin{figure}[ht!]
  \centering
  \includegraphics[width=0.9\textwidth]{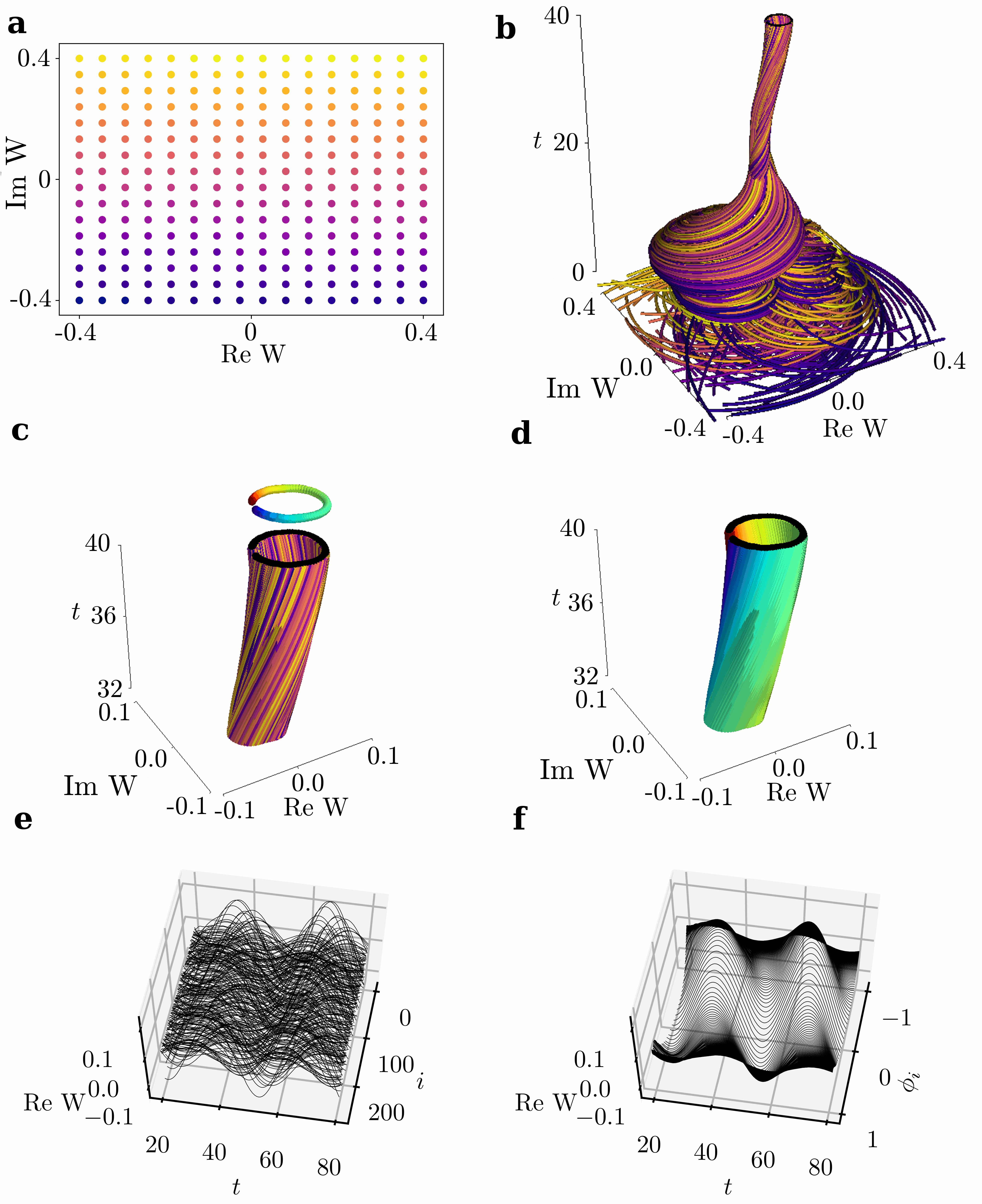}
  \caption{(a) Initial condition of the Stuart-Landau ensemble, Eq.~\eqref{eq:sle},
    colored with ascending imaginary part of $W_k$.
    (b) Trajectories obtained from integrating the initial conditions of (a) with the same
    color coding as in (a). The last snapshot is marked by black dots.
    (c) Zoom in to the upper part of (b), with the last snapshot marked by black dots.
    Above it, the last snapshot is color coded {\em based on the ordering of the oscillators along the curve at that moment}.
    (d)~Zoom in on the top part of (b), but now with the new color coding.
    (e) Trajectories of the real part of the $W_k$,
    arranged by their initial values $\mbox{Im} W$.
    (f) Trajectories of the real part of the $W_k$,
    arranged by the new color coding $\phi_i$ as in (d). (Finding $\phi_i$ is discussed in the text).}
  \label{fig:1}
\end{figure}

Suppose we initialize an ensemble of $N=256$ oscillators with values $W_k$ on a regular grid,
as shown in Fig.~\ref{fig:1}(a).
The color coding thereby correlates with the imaginary part of $W_k$.
Integrating this initial condition using Eq.~\eqref{eq:sle} with coupling constant $K=1.2$ and intrinsic frequencies $\omega_k$ distributed equally spaced within the interval $\left[-1.5, 1.9\right]$ yields
the dynamics in Fig.~\ref{fig:1}(b):
although the behavior appears quite irregular at the beginning, it quickly settles onto a cylinder-like structure.
Note that the color coding is still the same.
After the transients decay, the agents appear arranged on
this structure in an irregular manner if colored based on their initialization, see the zoom in of the upper part as shown in Fig.~\ref{fig:1}(c).
Using manifold learning, we will show that it is possible to find a
parametrization of the agents (a ``different coloring'') in which the dynamics appears more ordered and regular.
This is shown by the new color coding of the last snapshot in Fig.~\ref{fig:1}(c),
and the recolored attractor in Fig.~\ref{fig:1}(d).
Indeed, when contrasting the time series of the agents in the original color coding $i$
(Fig.~\ref{fig:1}(e)) and the new color coding $\phi_i$  (Fig.~\ref{fig:1}(f)),
we argue that the dynamics appear more regular in a space parametrized by $\phi_i$,
suggesting the possibility that the solution can be described by a PDE with $\phi_i$ and time as the independent variables.

The remainder of this article is organized as follows: First,
we illustrate our approach through a caricature, where we start with a known 
PDE in a predefined spatial variable. We observe the dynamics at a number of mesh points in this
known space, but then we ``scramble" the time series ourselves, on purpose, 
concealing the spatial coordinates of where the behavior was observed.
We obtain a predictive PDE description in a learned emergent ``spatial'' or ``heterogeneity'' coordinate $\phi_1$,
discovered through data mining these scrambled behaviors.
%
We then confirm that this emergent coordinate
is one-to-one with the (discarded) physical location $x$ of the original mesh points. 

Returning to our 
globally-coupled oscillator ensemble,
we show how to extract an intrinsic space coordinate,
and learn a PDE description in this parametrization and time.
We then study parametric dependencies of this PDE: we sample dynamics at parameter values bracketing a (collective) Hopf bifurcation.
Using this data, we show that learning a PDE with an additional input for a parameter can capture
the location and nature of bifurcations in this parameter.

We then go beyond a single ``emergent space'' dimension:
We vary the nature of the oscillator ensemble by adding a second parameter, beyond
the oscillator frequency.
Data mining discovers that the description of the agent behaviors is now two-dimensional.
We again learn a PDE describing the agent dynamics - now in two ``emergent space coordinates'' and time.

We conclude with a discussion of the approach and its shortcomings, and what we perceive as open questions and directions for future research. We also discuss the explainability of the learned emergent coordinate(s) for such agent-based systems.
Details on the algorithms and numerical methods are summarized in the Methods section.
The code to reproduce the results will be made available under \href{https://github.com/fkemeth/emergent_pdes}{{https://github.com/fkemeth/emergent\_pdes}} upon publication.

\section{Results}
\begin{figure}[ht]
  \centering
  \includegraphics[width=\textwidth]{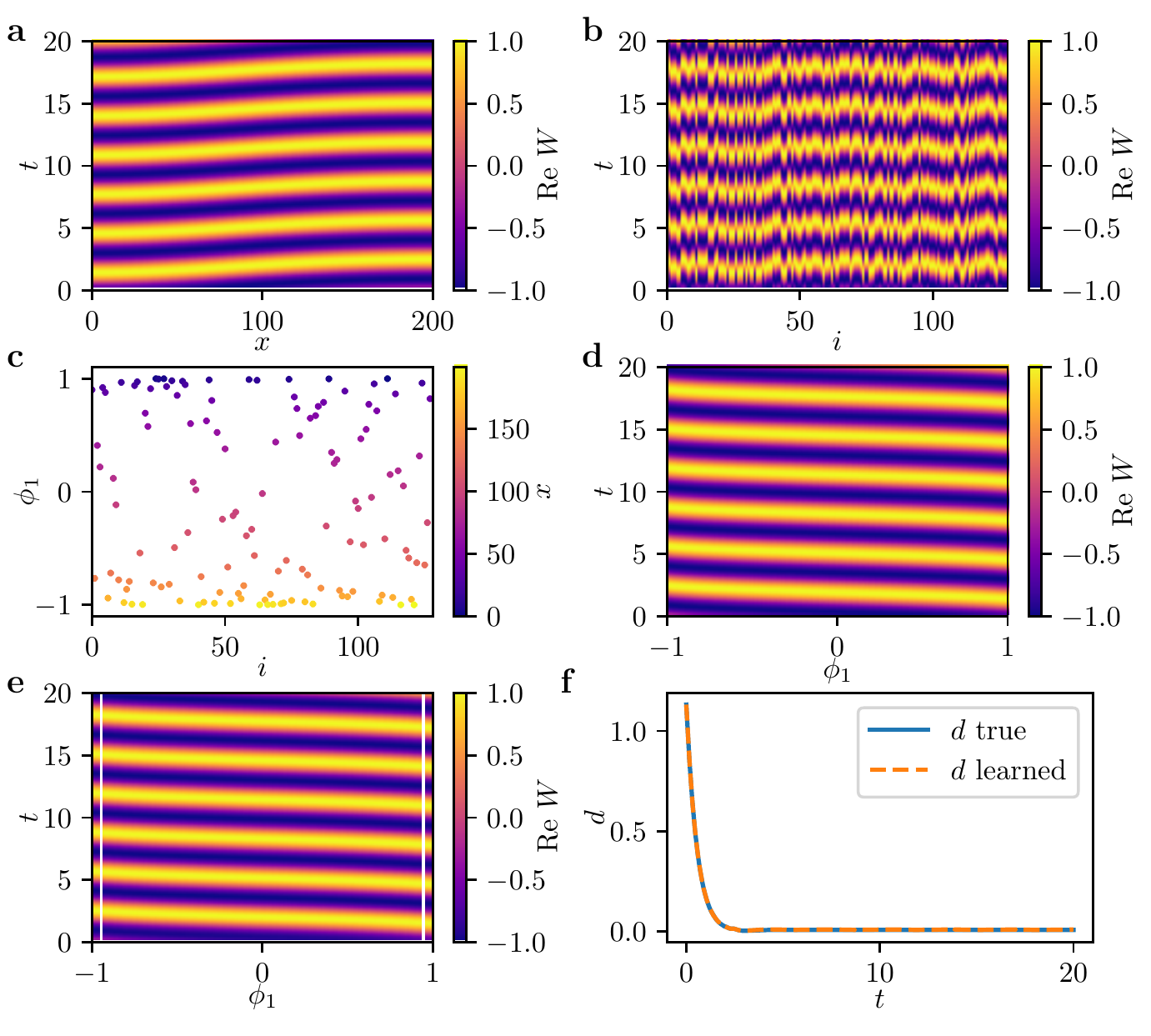}
  \caption{(a) The real part of the complex field $W(x, t)$
    obtained from simulating Eq.~\eqref{eq:cgle} with $N=128$ mesh points 
%
    after initial transients have decayed.
    (b) Removing the spatial label yields a collection of $N$ time series plotted here
    in random sequence.
    (c) Using manifold learning (here diffusion maps), one finds that there exists a one-dimensional
    parametrization $\phi_1$ of these time series. Each point corresponds to one of the $N$
    time series, and is colored by its actual spatial location $x$.
    (d) The real parts of the time series parametrized by $\phi_1$.
    (e) Real part of simulation predictions for the complex variable $W$ starting from an initial condition in our test set, using the partial differential equation model learned with $\phi_1$ as the spatial variable. 
    Since no analytical boundary conditions are available, we provide the true values near the boundaries during integration, within a ``corridor" indicated by white vertical lines.
    (f) Smallest Euclidean distance in $\mathbb{C}^N$ between the transients and the true attractor at each time step: true PDE (blue), learned PDE (orange).}
%
  \label{fig:3}
\end{figure}

{\bf Learning PDEs in Emergent Coordinates}.
For an illustrative caricature, we use a PDE with a known independent space variable, 
before returning to our coupled agent example.
%
Consider the 1D complex Ginzburg-Landau equation,
a PDE for the evolution of a complex field $W(x, t)$ in one spatial dimension $x\in\left[0, L\right]$, defined by
\begin{equation}
\frac{\partial}{\partial t} W(x, t) = W(x, t) + \left(1+ic_1\right)\frac{\partial^2}{\partial x^2} W(x, t) - \left(1-ic_2\right)|W(x, t)|^2 W(x, t)
\label{eq:cgle}
\end{equation}
with real parameters $c_1=1$, $c_2=2$, $L=200$ and, here, no-flux (Neumann) boundary conditions.
We integrate starting with initial condition
\begin{equation*}
  W(x, 0) = \frac{1+\cos{\frac{x\pi}{L}}}{2}
\end{equation*}
using a finite-difference method in space and an implicit Adams method for integration, and sample data after initial transients have decayed, i.e. after 4000 dimensionless time units.
This spatiotemporal evolution is depicted in Fig.~\ref{fig:3}(a).

The spatial coordinate $x$ is discretized into
$N=128$ equidistant points $x_k$.
Eq.~\eqref{eq:cgle} thus yields $N$ (here complex) time series $W_k(t)$ at each mesh point $x_k$.
We can think of the behavior at each mesh point as the behavior of an agent in an ensemble of interacting agents. 
Assuming the $x_k$ label of each agent is not available
(cf. Fig.~\ref{fig:3}(b), where the agents are parametrized by a random index $i$);
is it possible to find a collective description of the
dynamics in these time series based on a data-driven, ``emergent'' spatial variable,
and in the form of a partial differential equation, involving partial derivatives in this variable?

We accomplish this by extracting an intrinsic independent coordinate from the time series data.
As proposed in 
Ref.~\cite{kemeth18_emerg_space_distr_data_with}
we use diffusion maps (each of the scrambled time series is a data point) to extract a coordinate parametrizing
the ensemble of time series: the leading diffusion map component of each time series (of each data point); see Methods.
It may be qualitatively helpful (even though we use a nonlinear manifold learning algorithm) to think of this as performing principal component analysis (PCA) on the ensemble of time series (each of them is a data point) and then keepinging the leading PCA component as an emergent spatial coordinate.
This emergent coordinate is used to parametrize a useful embedding space in which to learn a PDE.

For the time series data in Fig.~\ref{fig:3}(b),
one indeed finds a one-dimensional parametrization of the $W_k$,
which is shown in Fig.~\ref{fig:3}(c).
This coordinate is one-to-one with the (original, ``forgotten'') spatial coordinate $x$
(see color coding in Fig.~\ref{fig:3}(c)).
Even not knowing knowing the spatial location of the mesh points,
we can still extract a data-driven parametrization $\phi_1$ and set out to learn a
PDE with this coordinate as the spatial dimension.
The data parametrized this way is depicted in Fig.~\ref{fig:3}(d).
Note that $\phi_1$ is one-to-one with, {\em but not identical}, to $x$.
In particular, it is also flipped (see the mirrored Figs.~\ref{fig:3}(a) and~\ref{fig:3}(d))
and slightly ``compressed'' due to edge effects at small and large~$x$.

We now set out to learn a PDE description based on partial derivatives in $\phi_1$,
\begin{equation}
  \frac{\partial}{\partial t} W(\phi_1, t) = f\left(W, \frac{\partial W}{\partial \phi_1},
  \frac{\partial^2W}{\partial \phi_1^2}, \frac{\partial^3W}{\partial \phi_1^3}\right)
\label{eq:pde}
\end{equation}
where $f$ is represented by a fully connected neural network.
See Methods for details on the neural network architecture and the data sampling.
A number of issues arise in learning such a PDE in $\phi_1$:
\begin{itemize}
\item Since $\phi_1$ is not identical to $x$, trajectories $W$ are not equally spaced.
  To calculate a finite difference approximation of $\partial^n W/\partial \phi_1^n$,
  we interpolate the $\phi_1$-parametrized data using cubic splines and sample $W$ at
  $N=128$ equidistant points on the interval $\left[-1, 1\right]$.
\item Due to the deformation of the space coordinate, the boundary conditions in the transformed variable may no longer be obvious.
  We therefore learn $f$ only in the interior of the $\phi_1$ domain. 
  When we simulate the learned PDE, we provide (as boundary conditions) a narrow space-time data corridor
  as needed. The imposition of such ``{\em finite corridor} boundary conditions'' will be particularly important for agent-based systems,
  where the form of effective boundary condition formulas (like Dirichlet, Neumann or Robin) in the emergent space is not known {\em a priori}. 
\item PDEs are infinite dimensional; we cannot sample the full state space, and so 
our learned surrogate PDE will ``not know'' the dynamics in all state space directions.
  Various techniques proposed in recent years (especially in imitation learning) attempt to regularize surrogate dynamical systems, .
  These include contraction theory~\cite{lohmiller98_contr_analy_non_linear_system,singh19_learn_stabil_nonlin_dynam_with,blocher17_learn,sindhwani18_learn_contr_vector_field_stabl_imitat_learn},
  and convex neural networks~\cite{brandon17_input_convex_neural_networks,manek20_learn_stabl_deep_dynam_model}. They rely on the existence of a Lyapunov function;
  other approaches include Jacobian
  regularization~\cite{hoffman19_robus_learn_with_jacob_regul,pan18_long_time_predic_model_nonlin}.
  However, they usually involve additional loss terms
  or are computationally expensive.\\
  Here, we instead regularize the output of the learned PDE as follows:
  First, we sample several transients close to the limit cycle solution of
  the complex Ginzburg-Landau equation (a ``tube'' in phase space). 
  Then, we create a truncated singular value decomposition (SVD) based on all the sampled transients.
  During inference, we filter the state obtained by integration of the neural network output by projecting it back onto this truncated SVD subspace, thus keeping the predicted trajectories there.
\end{itemize}
Integrating from an initial snapshot using the learned PDE $f$ in the emergent variable $\phi_1$ is shown in
Fig.~\ref{fig:3}(e).
Notice the close correspondence between predicted and actual dynamics,
cf. Fig.~\ref{fig:3}(d).
We also investigate whether nearby transients approaching the attractor are captured accurately
by the learned PDE.
To test this we integrate starting from an off-attractor snapshot using both the original PDE {\em and} the learned PDE,
and plot the smallest Euclidean distance in $\mathbb{C}^N$ between the transients obtained this way and the true attractor over time.
%
See Fig.~\ref{fig:3}(f) for a measure of the true distance (blue) and the distance when integrating with the learned model (orange).
There is good correspondence between the two curves, rendering the blue trajectory
barely visible.\\
In the next Section, we will follow the same approach, but now for a system where {\em there
is no original space coordinate}.

\begin{figure}[ht]
  \centering
  \includegraphics[width=\textwidth]{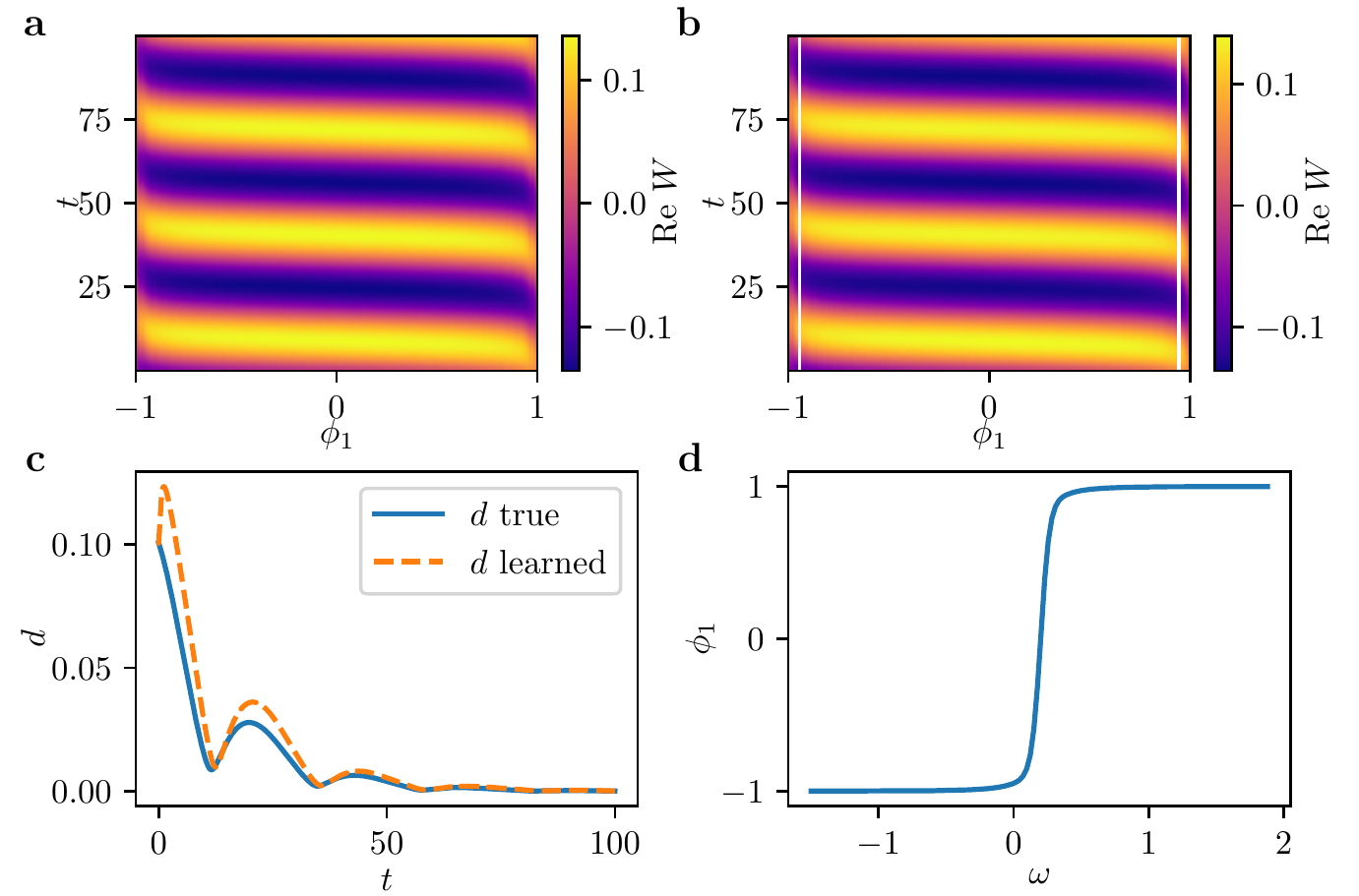}
  \caption{(a)~Real part of the complex variable $W$ for a system of $N=512$ oscillators,
    parametrized by the first emergent diffusion mode $\phi_1$.
    (b) Dynamics obtained from the learned model by integrating starting from the same initial snapshot as in (a).
    (c) Smallest Euclidean distance in $\mathbb{C}^N$ at each time step between the transients and the true attractor for the true PDE (blue) and the learned PDE (orange).
    (d) The first diffusion mode $\phi_1$ as a function of the intrinsic frequencies $\omega$ of the
  oscillator ensemble.}
  \label{fig:6}
\end{figure}

{\bf Learning Partial Differential Equations for Coupled Stuart-Landau Oscillator Dynamics}
Recall the original problem, Eq.~\eqref{eq:sle}, of an
ensemble of mean-coupled Stuart-Landau oscillators,
\begin{equation}
  \frac{\mathrm{d}}{\mathrm{d}t}  W_k = \left(1+i \omega_k\right) W_k - \left|W_k\right|^2W_k  + \frac{K}{N}\sum_j\left( W_j - W_k\right)
\end{equation}
with $k=1,\dots,N$ and the real coupling constant $K$.
The intrinsic frequencies $\omega_k$ are taken linearly spaced in the interval $\left[-\gamma+\omega_0, \gamma+\omega_0\right]$.
Depending on the parameters $K$ and $\gamma$, a plethora of different dynamical phenomena are known to arise.
Examples range from frequency locked oscillations and quasiperiodic dynamics to chaos and oscillator death.
See Ref~\cite{matthews90_phase_diagr_collec_behav_limit_cycle_oscil} for a more detailed discussion.
Here, we fix $K=1.2$, $\gamma=1.7$ and $\omega_0=0.2$ - resulting in periodic, synchronized oscillations:
the oscillators in the ensemble oscillate with a common frequency and
maintain a constant mutual phase difference.
The real part of such dynamics is depicted in Fig.~\ref{fig:6}(a),
parametrized by $\phi_1$, the first independent diffusion map mode.
%
As for the complex Ginzburg-Landau equation, we sample data not only on the attractor,
but also on transients in its neighborhood approaching it. 
These long-term dynamics can be thought of as lying on an attracting slow manifold; see Methods. 
The predictions from an initial condition on the limit cycle using the learned PDE model are depicted in Fig.~\ref{fig:6}(b),
and closely resemble the actual dynamics,
as depicted in Fig.~\ref{fig:6}(a).
The model also captures the dynamics approaching the limit cycle.
This can be visualized by integrating from initial conditions on the slow manifold
but off the attracting limit cycle.
We integrated such an initial condition from our test set using forward Euler and both the full ODE system,
Eq.\eqref{eq:sle}, as well as the learned emergent PDE model.
The smallest Euclidean distance in $\mathbb{C}^N$ between these transients and the true attractor at each time step is depicted in Fig.~\ref{fig:6}(c).
Note that both the true and learned transients converge to the limit cycle at a similar rate,
and the learned PDE trajectory approximates the behavior of the full ODE system well. 
%
In an attempt to obtain a physical meaning of the emergent coordinate $\phi_1$, 
we plot it as a function of the intrinsic
frequency $\omega$ of the oscillators in Fig.~\ref{fig:6}(d).
It becomes obvious that the two quantities are one-to-one, analogous to the ($\phi_1$, $x$) pair in
the complex Ginzburg-Landau example above: our data mining has ``discovered" the heterogeneity of the ensemble, and uses it to parametrize the dynamics.
%
Knowing the equations and how $\omega_k$ enters in them, one could analytically attempt to derive Ott-Antonsen-type equations (for phase oscillators) in $\omega$ space~\cite{Ott-Antonsen}. We know neither the equations, nor the $\omega_k$ (and the oscillators are not phase oscillators to boot); everything here is data-driven. 

\begin{figure}[ht]
  \centering
  \includegraphics[width=\textwidth]{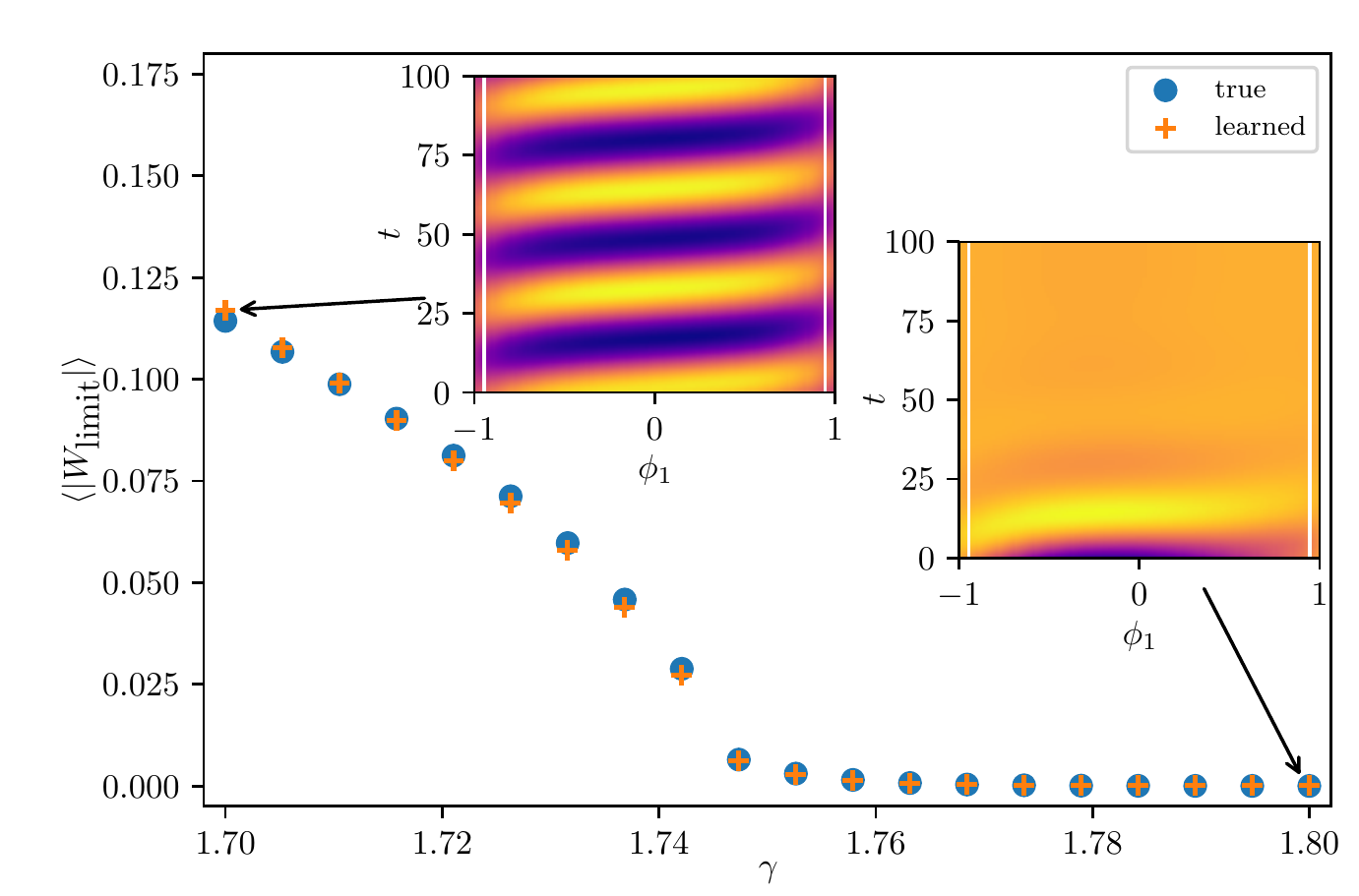}
  \caption{
    Computational bifurcation diagram by plotting the mean amplitude $\langle | W_{\mbox{limit}}| \rangle$
    averaged over the ensemble at the limit set.
    In particular, we integrate from random initial conditions close to the limit set for $T=10000$
    dimensionless time units for the Stuart-Landau ensemble (blue circles) and the learned PDE
    (orange crosses). A mean amplitude indicates convergence to the fixed-point $W=0$ $\forall \omega$,
    whereas a non-zero $\langle | W_{\mbox{limit}}| \rangle$ indicates oscillations with
    finite amplitude.
    The color codings of the insets show real part of the complex variable $W$ obtained from
    integrating an initial condition close to the fixed point $W_k=0$ with $\gamma=1.8$ (right inset) and close to the limit cycle with $\gamma=1.7$ (left inset)
    using the learned model and employing explicit forward Euler for $\gamma=1.8 > \gamma_H$.}
  \label{fig:7}
\end{figure}

Having been successful in capturing the attractor and its nearby dynamics for a single
parameter value, it becomes natural to explore whether the learned PDE can also capture bifurcations: qualitative changes in the dynamics when changing system parameters.
In particular, for $\gamma=\gamma_H\approx 1.75$,
the Stuart-Landau ensemble undergoes a collective Hopf bifurcation, at which the amplitude of the
oscillations shown in Fig.~\ref{fig:6} vanishes.
For $\gamma>\gamma_H$, a stable fixed point ensues, in which
all individual amplitudes of the respective oscillators are zero, also called oscillator death~\cite{aronson90_amplit_respon_coupl_oscil}.
We now collect data for training at several $\gamma$ values, linearly spaced in the interval $\left[1.7, 1.8\right]$,
on both sides of the Hopf bifurcation; the $\gamma$ value was provided
as additional input to the model.
We again perturbed along the slow stable eigendirections of each attractor, see Methods, collecting transients that inform the model about nearby dynamics.
We then learned a PDE of the form
\begin{equation}
  \frac{\partial}{\partial t} W(\phi_1, t) = f\left(W, \frac{\partial W}{\partial \phi_1},
  \frac{\partial^2W}{\partial \phi_1^2}, \frac{\partial^3W}{\partial \phi_1^3}; \gamma\right).
\label{eq:pde_alpha}
\end{equation}
The learned dynamics, starting from an initial oscillator ensemble profile, and integrated using the learned model
are shown in Fig.~\ref{fig:7} for $\gamma < \gamma_H$ (left inset) and for $\gamma > \gamma_H$ (right inset).
We observe the transient dynamics approaching the fixed point $W=0 \, \forall \omega$
%
for $\gamma = 1.8$, as the true dynamics (not shown here) also do. 
%
%
Validating the approach further, we start at random initial conditions in the slow eigenspace of the attractor at different $\gamma$ values
using the Stuart-Landau system, Eq.~\eqref{eq:sle}, as well as the learned PDE model.
For both models, we record a snapshot after $T=10000$ dimensionless time units
and calculate its average amplitude $\langle | W_{\mbox{limit}}| \rangle$.
An average amplitude equal to zero then indicates that the initial condition converged to the
fixed point $W=0 \, \forall \omega$ under the respective model, whereas a nonzero amplitude
indicates convergence to the (collective/spatiotemporal) limit cycle.\\
The resulting $\langle | W_{\mbox{limit}}| \rangle$ values for different $\gamma$ are
shown in Fig.~\ref{fig:7}, with blue circles for the original dynamics and orange crosses for
the learned dynamics.
The Hopf bifurcation manifests itself in the sudden increase in amplitude
when $\gamma$ is varied.
Note the close correspondence between the learned model and the original oscillator
system: both converge to a fixed point for $\gamma > \gamma_H \approx 1.75$, and to the
limit cycle for $\gamma < \gamma_H \approx 1.75$.
\begin{figure}[ht]
  \centering
  \includegraphics[width=0.9\textwidth]{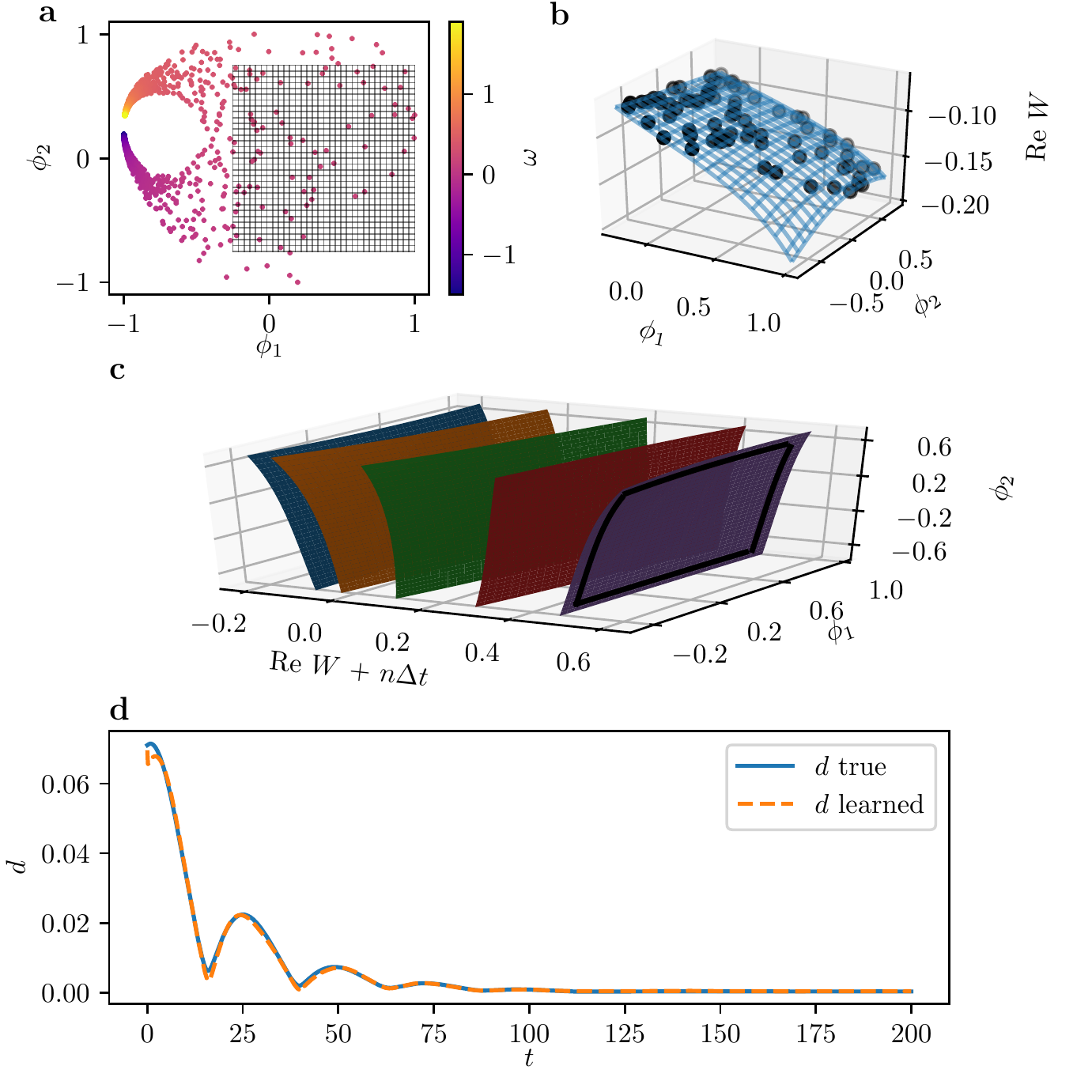}
  \caption{(a) Diffusion maps embedding of an ensemble of Stuart-Landau oscillators with
  two heterogeneous parameters. The color thereby encodes one of the two parameters, the intrinsic frequencies $\omega$. The grid indicates the space in which we resample the data and learn a PDE description. (b) Data points and fit of the real parts of $W$ in the grid shown in (a).
  (c) Snapshots of the real parts of $W$ obtained using the learned model at different time points
  $\Delta t = 4$ apart. The black lines in the last snapshot indicate the ``boundary corridors" provided to the model
  during integration.
  (d) Smallest Euclidean distance in $\mathbb{C}^N$ between the transients and the true attractor, obtained by using the true PDE (blue, Eq.~\eqref{eq:sle_2d}) and the learned PDE (orange, Eq.~\eqref{eq:pde_2d}) for integration.}
  \label{fig:8}
\end{figure}

{\bf Two emergent spatial coordinates}.
The approach can easily be extended to situations with more than one emergent spatial dimension,
that is, problems in which more than one diffusion map component become necessary to parametrize the inherent heterogeneity of agent behaviors.
As an example, consider our Stuart-Landau ensemble above, but now with {\em two heterogeneous parameters}, $\omega_k$ and $\lambda_k$,
\begin{equation}
  \frac{\mathrm{d}}{\mathrm{d}t} W_k = \left(\lambda_k+i \omega_k\right) W_k - \left|W_k\right|^2W_k  + \frac{K}{N}\sum_{j=1}^N\left( W_j - W_k\right).
  \label{eq:sle_2d}
\end{equation}
The $\omega_k$ are taken linearly spaced in the interval $\left[-\gamma+\omega_0, \gamma+\omega_0\right]$, 
while the $\lambda_k$ are drawn uniformly from the interval $\left[0.9, 1.1\right]$.
Using diffusion maps, one confirms that there is a two-parameter family of oscillator behaviors.
The two independent modes $\phi_1$ and $\phi_2$ are shown in Fig.~\ref{fig:8}(a),
color-coded by the $\omega_k$.
In the following, we set out to learn a PDE in this two-dimensional emergent space spanned
by $\phi_1$ and $\phi_2$.
In order to do so, we interpolate the available data on the rectangular grid shown in Figs.~\ref{fig:8}(a-c);
both the real and the imaginary components of the $W_k$ are interpolated at equidistant points.
A snapshot of this is shown in Fig.~\ref{fig:8}(b).
As earlier, we sample transients initialized on the attractor and in the slow manifold, but now learn a PDE $f$ of the form
\begin{equation}
  \frac{\partial}{\partial t} W(\phi_1, \phi_2, t) = f\left(W, \frac{\partial W}{\partial \phi_1}, \frac{\partial W}{\partial \phi_2},
  \frac{\partial^2W}{\partial \phi_1^2},\frac{\partial^2W}{\partial \phi_2^2}\right).
\label{eq:pde_2d}
\end{equation}
To evaluate the quality of the learned model, we integrate starting with initial snapshots (approximately) on the slow manifold but off the attractor, both using the learned model and the original system, and plot the closest distance between the two transients and the true attractor as a function of time.
This is shown in Fig.~\ref{fig:8}(d). Again, the learned model captures the transient dynamics approaching, as well as along, the attractor.

\section{Discussion}
We have seen that it is possible to learn a predictive model for the dynamics of coupled agents based on
local partial derivatives with respect to one (or more) emergent, data-driven ``spatial variable(s)'' and time, that is, in the form of a partial differential equation.
As an example, we investigated an ensemble of mean-coupled Stuart-Landau oscillators, where each oscillator has an intrinsic frequency $\omega_k$.
Using manifold learning (here diffusion maps), we were able to extract an intrinsic coordinate $\phi_1$ from time
series segments of these oscillators.
Starting with just a single parameter value $\gamma=1.7<\gamma_H$,
our results indicate that a model based on a few partial derivatives with respect to $\phi_1$
is able to accurately capture the collective dynamics in the slow manifold and on the final attracting limit cycle.
These results extend to the case in which data is sampled for different $\gamma$ values on both sides
of the Hopf bifurcation point $\gamma_H$.
The learned PDE then modeled successfully the slow transients towards
either the stable limit cycle or the stable fixed point, depending on the parameter.
We then extended our analysis to the case where the oscillators depend
on {\em two heterogeneity parameters},
and the corresponding diffusion maps embedding is two-dimensional.
This then results in a PDE in a two-dimensional emergent space.

For a successful implementation of our approach we employed a systematic way of sampling
training data:
From a given limit set, we perturb along the slow stable manifold, and sample transients approaching the attractor.
This sampling strategy is assisted by estimates of the slow stable directions (and their time scales) through the linearized system Jacobian, that help produce informative initial conditions.
Because of the ``fast-slow'' nature of the dynamics, we found that
starting practically anywhere and integrating for a short time
will bring the dynamics close to this slow manifold.\\
This ought to also be the case when collecting experimental data (discarding short initial transients to the
slow manifold). 
Clearly, the model cannot be expected to learn the right asymptotic behavior in dimensions in which it has seen no data. 
This can lead to instabilities when attempting to predict the long term dynamics
of the system.
We addressed this problem through filtering, in particular through a truncated SVD regularization.
An SVD basis was constructed from the training data, and, during inference,
we filtered by projecting the predictions on this basis; the predicted dynamics cannot leave the space spanned by the truncated SVD.
This introduces an additional hyperparameter to the model: the dimension after which to truncate the SVD used for filtering.
Too many dimensions may allow for instability in the predictions (lack of training data);
too few leads to poor representations and distorted dynamics. 
Our threshold was empirically chosen by trial and error; the subject is worthy of a more detailed study.

An important question in deciding which PDE model to learn, is how many ``emergent spatial'' derivatives one has to include in the PDE right hand side.
In other words, how can one decide when $\partial W/\partial t$ is well approximated by
$W$ and its derivatives with respect to $\phi_1$?
For Gaussian process regression, recent work using Automatic
Relevance Determination helps tackle this problem~\cite{lee20_coars_scale_pdes_from_fine}.
In our case we again decided empirically, by trial and error; a more thorough study must clearly follow.
In addition, the issue of boundary conditions in emergent space (here we used narrow ``boundary corridors"), as well as what
constitutes a well posed problem for an operator identified in a data-driven way constitute important (and challenging) questions to pursue; we mention here the possibility of using the approach of the ``baby-bathwater'' scheme in \cite{li07_decid_natur_coars_equat_throug_micros_simul}.

Fig.~\ref{fig:7}(b) indicates that the learned model
captures qualitative changes in the dynamics when changing a system parameter,
here a Hopf bifurcation from a fixed point for $\gamma>\gamma_H$ to collective oscillations
for $\gamma < \gamma_H$.
More quantitatively, we reported
the leading spectrum of the linearization of the model evaluated at the fixed point.
This was obtained using automatic differentiation of the neural network model with respect to 
its inputs. 
Such computations can shed more light on the similarities and differences of
agent-based simulations and their emergent PDE descriptions.
In this paper we focused on a particular regime in parameter space. 
However, our approach can easily be extended to more intricate dynamics that are known
in such a Stuart-Landau ensemble; informative examples are included in the videos \href{https://github.com/fkemeth/Emergent_PDE_Videos/blob/master/emergent_pde1.avi}{SI1} and \href{https://github.com/fkemeth/Emergent_PDE_Videos/blob/master/emergent_pde1.avi}{SI2}.

Historically, it is known that physical phenomena modeled at the fine scale through
atomistic/stochastic/agent-based simulations are often
well approximated using closed partial differential equations in terms of a few of
their collective observables (e.g. moments of the particle distribution, such as the agent density).
Our approach will be useful when we believe that such effective, collective PDE models in principle exist, but the closures required to write them down are not known.
It can also provide useful results in regimes where the strong mathematical assumptions required to provably obtain explicit closures can be relaxed.
This is an area where equation-free multiscale numerics has been used to solve the equations without writing them down, and where manifold learning has been used to even perform this solution ``(dependent) variable free'', that is, in terms of dependent variables not known a priori, but revealed through data mining of detailed simulations (see, for example, the discussion in~\cite{erban07_variab_free_explor_stoch_model}).
All scientific computation in latent space (e.g. see~\cite{chiavazzo14_reduc_model_chemic_kinet_via} and~\cite{lee20_model_reduc_dynam_system_nonlin}) falls in this class.

What is different and exciting (to us at least) in the present study, is the extension of this
approach to problems where there are no obvious {\em independent spatial variables} - dynamics of coupled oscillators, dynamics on and of networks, dynamics of ``systems of interacting systems'', where ``the right space'' for modeling the problem is not known {\em a priori}. 
Writing models in such an emergent ``activity space'', with emergent space {\em and even emergent time(!)}~\cite{kemeth18_emerg_space_distr_data_with}
coordinates may become a useful method for the modeler: a tool that extends the toolkit for linking domain science knowledge at the detailed level with machine/manifold learning to build useful, predictive models.

Here, we chose a model based on local descriptors, {\em local in the emergent space}.
One can speculate about contexts in which such a local description might be beneficial. It certainly is more humanly parsimonious/compact to write down than the detailed list of all units and all interactions. It may also be convenient if one needs to make predictions with limited memory
(limited ``fast cpu memory'' so to speak). We do not need to know what every unit is doing - we look at the activity of similar units (that are already embedded nearby in emergent space) and make predictions based on smoothness (mathematically expressed through Taylor series) and the behavior of the neighbors. 
Our emergent space can then be thought of as {\em a space where nearby (observations of) behaviors come already usefully clustered}.  Alternatively, we can think of this space as embodying a useful ``attention geometry'' - the behaviors we need to pay attention to (because of their similarity) in order to make a prediction, are already our neighbors in this space. Geometric proximity in the emergent space saves us then from having to search for comparable behavior histories across all interacting units in physical space-time. This enables us to exploit smoothness across behavior histories in order to make local predictions with only a few nearby data.

We touched briefly upon the explainability of our emergent spatial coordinates by showing that our $\phi_1$ was one-to-one with, and thus calibratable to, the oscillator intrinsic frequencies - the agent heterogeneity.
The suggested approach then is to (a) decide how many emergent independent variables are necessary; (b) ask a domain scientist for physical quantities that may ``explain them'' and then (c) to test whether the explainable and the data-driven parametrizations
are one-to-one on the data (the determinant of the Jacobian of the transformation is bi-Lipschitz, bounded away from zero and from infinity, on the data, e.g.~\cite{sonday09_coars_grain_dynam_driven_inter,frewen10_coars_collec_dynam_animal_group,meila2018regression}).

Clearly, the explainability of predictive, generative equations in terms of data-driven dependent and independent variables, and operators approximated through machine learning is a crucial endeavor - when and why will we decide we trust results when we ``understand'' the algorithms, but do not ``understand'' the mechanistic, physical steps underlying the observations of what we model?
Will a ``different understanding" arise in latent/emergent space -analogous, say, to describing operators
in Fourier space rather than physical space, or studying control in Laplace space rather than state space?
From flocking starlings to interacting UAV swarms, this promises to be an exciting playing field for contemporary modelers. 

\section{Methods}
{\bf Diffusion Maps}
Diffusion maps use a kernel function to weigh pairwise distances between data points
~\cite{Coifman2006,Nadler2006}, typically the Gaussian kernel
\begin{equation*}
  k(x, y) = \exp\left(- \frac{\lVert x - y \rVert^2}{\epsilon}\right)
\end{equation*}
with a predefined kernel scale $\epsilon$ and a Euclidean distance metric, which we adopt here.
The data points $x, y$ are, in our case, the $N$ time series (each of length $8\cdot 10^5$; see below), resulting in a
$K \in \mathbb{R}^{N\times N}$ kernel matrix.
Row-normalizing this kernel matrix yields a Markov transition matrix,
also called diffusion matrix, and its leading independent eigenvectors
corresponding to the largest eigenvalues can be used to parametrize the data~\cite{dsilva18_parsim_repres_nonlin_dynam_system}.

{\bf Example: Complex Ginzburg-Landau Equation}.
Consider the complex Ginzburg-Landau equation
\[\frac{\partial}{\partial t} W(x, t) = W(x, t) + \left(1+ic_1\right)\frac{\partial^2}{\partial x^2} W(x, t) -
\left(1-ic_2\right)|W(x, t)|^2 W(x, t)\]
in one spatial dimension \(x\), in a domain of length \(L\).
We solve this equation using the initial condition
\[W(x, 0) = \left(1+\cos \frac{\pi x}{L}\right)/2,\]
with zero-flux boundary conditions and parameter values \(c_1=1\), \(c_2=2\) and \(L=200\).
Numerically, we integrate using a three point stencil for the finite difference approximation
of the second derivative \(\partial^2/\partial x^2\) with \(N_{\mbox{int}}=256\)
discretization points and an implicit Adams method with \(dt=10^{-3}\) for the temporal evolution.
The resulting behavior is depicted in Fig.~\ref{fig:3}(a).
Data for training our model is sampled as described in the following:
For the number of training examples, we set \(n_{\mbox{train}}=20\) and for the number of test
examples \(n_{\mbox{train}}=1\), yielding \(n_{\mbox{total}}=21\).
At \(n_{\mbox{total}}=21\) points along the limit cycle shown in Fig.~\ref{fig:3}(a),
we sample data as follows:
At \(t_i = t_{\mbox{min}}=2000 + i d\tau\) with \(i\in\{0, \dots, n_{\mbox{total}}-1\}\) we perturb
the limit cycle by scaling the respective snapshot at \(t_i\) as \(0.9 \cdot W(x, t_i)\) and
\(1.1 \cdot W(x, t_i)\). We integrate both of these snapshots forward in time for \(T=20\) time units,
and sample data after each \(dt=10^{-3}\).
This results in two transients, each comprised of \(20001\) snapshots at each \(t_i\).
This means, in total there are \(2 \times 20000 \times 20 = 8\cdot 10^5\) snapshot data pairs for training,
and \(2 \times 20000\) for validation.
We subsequently downsample the data to $N=128$ points per snapshot.
In order to find a parametrization for the discretization points of the PDE, 
we concatenate the training time series of the \(N=128\) points, 
resulting in \(2\times 20000 \times 20\) long trajectories.
Then, we use diffusion maps with an Euclidean distance and a Gaussian kernel, and take the kernel scale \(\epsilon\) as the median of all squared distances.
This results in the one-dimensional parametrization $\phi_1$, as shown in Fig.~\ref{fig:3}(c).
We resample data on a regular grid in the interval $\left[-1, 1\right]$ using a cubic spline.
We estimate the time derivative at each point using finite differences in time,
\begin{equation}
\frac{\partial}{\partial t} W(x, t_j) = \partial_t W(x, t_j) \approx (W(x, t_j+dt)-W(x, t_j))/dt,
\label{eq:fd}
\end{equation}
yielding 20000 $(W(x, t_j)$, $\partial_t W(x, t_j))$ pairs per transient and $t_i$.\\
Using the $(W(x, t_j)$, $\partial_t W(x, t_j))$ pairs, we train a neural network \(f\) such that
\[ \partial_t W(x, t_j) \approx f(W(x, t_j))\]
in a supervised manner as follows:
We take \(N=128\) discretization points on each snapshot.
At these points we calculate the first \(n_{\mbox{derivs}}=3\) spatial derivatives
using a finite difference stencil of length \(l=9\) and the respective finite difference kernel for
each spatial derivative of the highest accuracy order that fits into \(l=9\).
The model thus takes the form
\[ \partial_t W(x_i, t_j) \approx f(W(x_i, t_j), \partial_x W(x_i, t_j), \partial_{xx} W(x_i, t_j),
\partial_{xxx} W(x_i, t_j))\]
with the derivatives calculated as described above.
Note that \(W(x, t)\) is complex, which means at each \((x_i, t_j)\) the input to the neural network
is $8$-dimensional for \(n_{\mbox{derivs}}=3\).
The network itself is composed of \(4\) fully connected hidden layers
with \(96\) neurons each and \(\mbox{tanh}\)
activation function (resulting in $\approx 28\cdot 10^3$ trainable parameters). The output layer contains two neurons with no activation function,
one neuron for the real and imaginary part of \(\partial_t W\), respectively.
The network weights are initialized uniformly using PyTorch's default weight initialization~\cite{PyTorch},
and are optimized using the Adam optimizer~\cite{kingma2017adam} with initial learning rate of \(10^{-3}\) and
batch size of 1024.
Mean-squared error between the predicted and actual $\partial_t W(x_i, t_j)$, Eq.~\eqref{eq:fd}, is taken as the loss.
The model is trained for 60 epochs, and the learning rate reduced by a factor of 2 if the
validation loss does not decrease for 7 epochs.
Needless to say, other general purpose approaches to learning the right-hand-side of the operator (Gaussian Processes~\cite{lee20_coars_scale_pdes_from_fine}, Geometric Harmonics~\cite{coifman06_geomet_harmon}, etc.) can also be used.

Inference is done by taking an initial snapshot of the validation data or on the limit cycle and integrating it forward in time using the learned model and an integration scheme such as forward Euler. 
At each time step, the boundary conditions (in the form of narrow boundary corridors) are taken from the ground-truth data.
%
The issue arises of the right width for these corridors, and, more generally, the prescription of boundary/initial/internal conditions appropriate for the well-posedness of the overall problem,
especially since the operator (the right hand side of the PDE) comes in the form of a ``black box''. This is already the subject of extensive research that we, among others, are pursuing~\cite{wellposedness}.

In addition, each predicted snapshot from the model is filtered as described in the following.
On the whole training data set, an SVD is performed.
Using the obtained \(U\) and \(V\) matrices, we can decompose each predicted snapshot during inference.
In doing so, we truncate the SVD decomposition after two dimensions, and reconstruct the snapshot.
This means that each snapshot is projected onto the two-dimensional subspace in which the training
data lives, and thus prevents directions that have not been sampled from growing during inference.
The resulting dynamics obtained from the learned model and using an initial snapshot on
the limit cycle is depicted in Fig.~\ref{fig:3}(e).
$4$-point wide boundaries are provided on both sides of the domain.
The learned dynamics can be investigated more clearly by comparing the true and the learned
transient dynamics towards the limit cycle.
To do so, we integrate a snapshot perturbed away from the limit cycle using the
complex Ginzburg-Landau equation and the learned model,
and calculate the smallest Euclidean distance in $\mathbb{C}^N$ at each time step of the
obtained trajectories to the limit cycle.
The results are shown in Fig. \ref{fig:3}(f).\\
We also carefully checked that the learned model is converged with respect to the number of discretization points $N$.

{\bf Example: Stuart-Landau Ensemble}
The dynamics as depicted in Figs.~\ref{fig:1} and~\ref{fig:6} are globally stable for the parameters considered here~\cite{matthews90_phase_diagr_collec_behav_limit_cycle_oscil}.
In fact, arbitrary initial conditions decay to the limit cycle exponentially.
Such behavior can be investigated in more detail using Floquet theory:
the convergence to the limit cycle can then be described by Floquet multipliers with their
associate eigendirections.
Since the limit cycle described above is stable, the absolute values of the Floquet multipliers
are less than one, except for one of them which equals one.
In particular, multipliers with large magnitude indicate slow attracting directions,
whereas multipliers with absolute values close to zero indicate fast decaying directions.
If both small and large Floquet multipliers are present,
then there exist transients with multiple time scales.
\begin{figure}[ht]
  \centering
  \includegraphics{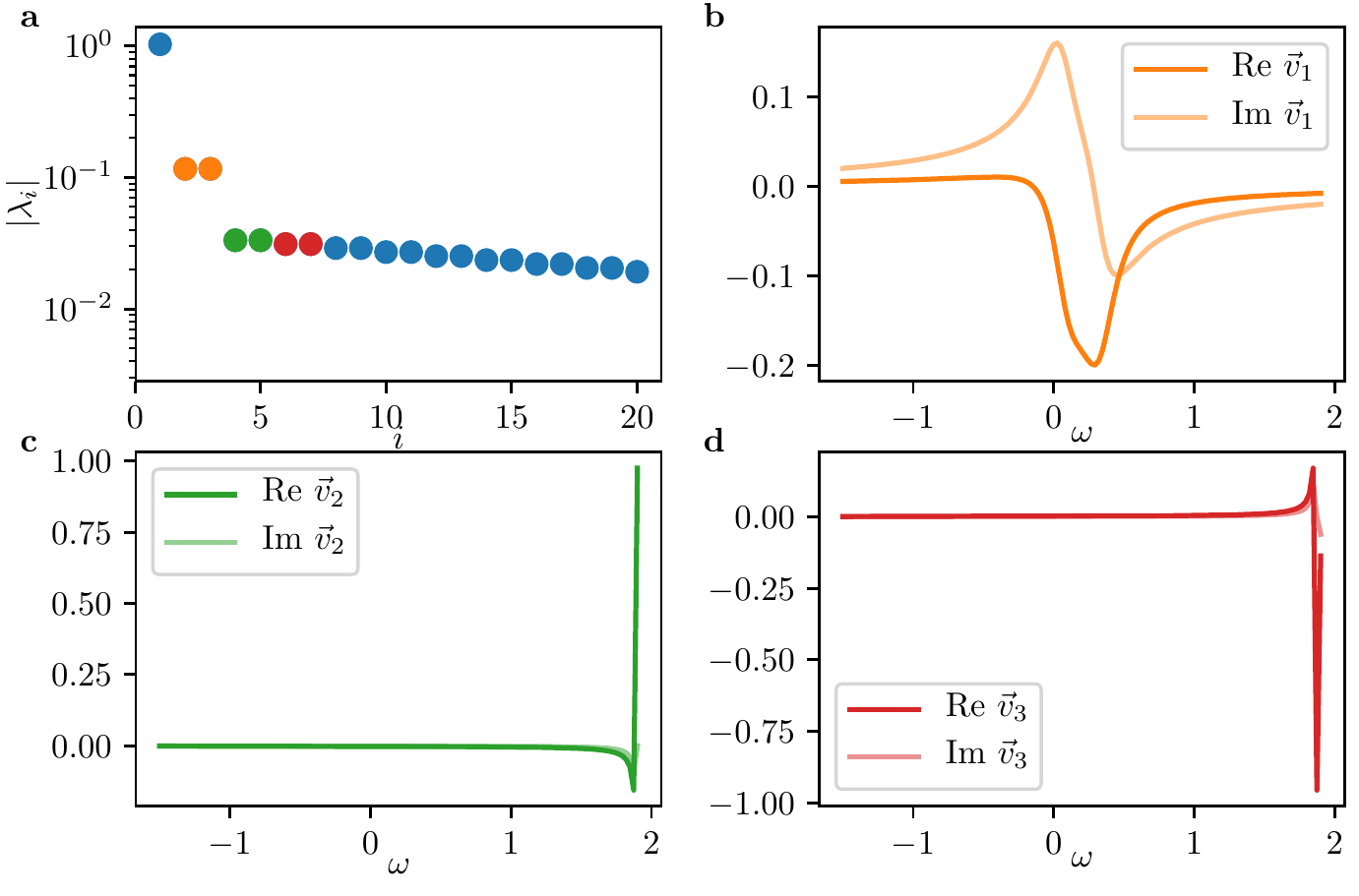}
  \caption{(a)~Floquet multipliers obtained from the monodromy matrix
    for the dynamics shown in Fig.~\ref{fig:6}.
    (b)~Eigendirection corresponding to the pair of complex conjugate multipliers
    $\lambda_2$ and$\lambda_3$ (marked in orange) indicating a slow attracting direction.
    (c, d) Eigendirections corresponding to the pairs of complex conjugate multipliers marked
    as green and red, indicating fast contracting directions.
    Note that since the $W_k$ are complex, the directions $\vec{v}_i$ are complex,
    with the real parts indicated as solid curves,
    and the imaginary parts indicated as shaded curves.}
  \label{fig:a1}
\end{figure}
Following Ref.~\cite{taylor92_coupled_double_toil},
we calculate the Floquet multipliers by calculating the monodromy matrix $\mathbf{V}$ along the limit cycle.
In particular, we obtain $\mathbf{V}$ by the integration
\begin{equation}
  \mathbf{V}(T) = \int_{t=0}^{t=T}\left.\frac{\partial F}{\partial x}\right|_{x(t)} \cdot \mathbf{V}\, \mathrm{d}t
  \label{eq:monodromy}
\end{equation}
with $\mathbf{V}(0) = \mathbf{I}_{2N\times 2N}$, $\mathbf{I}$ being the identity matrix,
and $T$ being the period of one oscillation.
The matrix $\frac{\partial F}{\partial x}$ represents the Jacobian of Eq.~\eqref{eq:sle} obtained analytically through differentiation and evaluated along the limit cycle.
The eigenvalues of $\mathbf{V}(T)$ then correspond to the Floquet multipliers,
with the corresponding eigenvectors being their respective directions.\\
The largest multipliers obtained this way, together with the three slowest eigendirections,
are depicted in Fig.~\ref{fig:a1}.
Notice the single multiplier equal to one represents the neutral direction along the limit cycle.
In addition, there is a pair of complex conjugate eigenvalues $\lambda_{2,3}\approx-0.4\pm 0.4i$ (orange in Fig.~\ref{fig:a1}).
Due to the magnitude of their real parts, the dynamics in this eigenspace is slow compared to the
subsequent eigendirections.
These eigendirections are, as apparent from Fig.~\ref{fig:a1}(b) smooth functions
of the frequencies $\omega_k$.
In addition, perturbations in this two-dimensional eigenspace spiral towards the stable limit cycle.\\
The directions of the subsequent multipliers affect only isolated oscillators.
In particular, the subsequent direction (green in Fig.~\ref{fig:a1})
following the slow eigenspace affects only the fastest oscillator,
that is, the oscillator with the largest intrinsic frequency $\omega_k$.
The next direction then perturbs the second fastest oscillator (red in Fig.~\ref{fig:a1}),
and so on.
The step-like structure of the Floquet multipliers highlights the multi-scale behavior of the coupled
oscillator system: The oscillation and the inward spiraling slow dynamics on one scale,
and the single oscillator dynamics towards the limit on the other, the fast scale.
These eigendirections with support on the ``most different'' oscillator are indicative of the
SNIPER bifurcation marking the ``edge of synchronization''.

We sample data by integrating system Eq.~\eqref{eq:sle} from random initial conditions, until the
dynamics are settled on the limit cycle. For $n_{lc}$ different points along the limit cycle,
we calculate the monodromy matrix from Eq.~\eqref{eq:monodromy} and estimate the least stable eigendirection $\vec{v}_1$ transverse to the limit cycle, presumably lying on the slow stable manifold. 
Then, we perturb in this direction by perturbing each point $W_{lc}$ on the limit cycle as $W_{lc}\pm \epsilon \vec{v}_1$, with $\epsilon=0.1$. This yields three initial points; integrating these points for a fixed amount of time then returns two transients towards the limit cycle and one trajectory on the attractor.
Here, we choose $n_{lc}=20$ for the training data, and $n_{lc}=5$ for the test data, and a time window of $T=200$ dimensionless time units with a sampling rate of $dt=0.05$, yielding $4000$ data points per trajectory, or $3 \cdot n_{cl}\cdot T/dt=240,000$ training data points and $60,000$ test data points.
The concatenated time series of length $3 \cdot n_{lc} \cdot T/dt$ then serve as input data points for diffusion maps; the possibility of using time series ``snippets" of different durations is explored in~\cite{kemeth18_emerg_space_distr_data_with}.
The temporal derivative $\partial_t W$ is then estimated using finite differences, cf. Eq.~\eqref{eq:fd}.
When also changing the system parameter $\gamma$ we provide for each data point the corresponding $\gamma$ value as additional input to the network.
In addition, the training data consists of uniform $\gamma$ values in $\left[1.7, 1.8\right]$, and the test data of randomly sampled $\gamma$ different from the training data.
In addition, we estimate an SVD basis from the complete training data. During inference,
the prediction of $f$ are reconstructed using this basis and a truncation with $n_s=3$
dimensions.\\
For the extraction of diffusion modes, we use a kernel scale of $\epsilon=20$ for the case when $\gamma$ is fixed and $\epsilon=10$ when we sample data with different $\gamma$ values.
Other hyperparameters and the model architecture are as described in the previous section.\\
For the case with two heterogeneous parameters, we simulate an ensemble of 
$N=2048$ oscillators and use \(n_{\mbox{train}}=5\) and \(n_{\mbox{train}}=1\).
We resample the data on a rectangular $32\times 32$ grid, as shown in Fig.~\ref{fig:8}(a) in diffusion maps space. Here, we use up to 2 derivatives in each dimension, and a finite difference kernel size of $l=5$. We thus provide boundaries of width 2 along the edges.
The model has 3 layers with 64 neurons each  (resulting in $\approx 9.3\cdot 10^3$ trainable parameters), and is trained for 20 epochs.

\textbf{Code Availability}
The code to reproduce the results will be made available under \href{https://github.com/fkemeth/emergent_pdes}{{https://github.com/fkemeth/emergent\_pdes}} upon publication.

\textbf{Acknowledgements} This work was partially supported by U.S. Army Research Office (through a MURI program), DARPA, and the U.S. Department of Energy. 

\textbf{Author Contributions} IGK. conceived the research which was planned jointly with all the authors. FPK performed a large part of the research, with contributions from TB, TT, FD, SJM and CRL. FPK and IGK initially wrote the manuscript, which was edited in final form with contributions from all the authors.

\textbf{Competing interests}
The authors declare no competing interests.


\bibliography{lit.bib}
\bibliographystyle{unsrt}

\end{document}